# Anisotropic geometrodynamics in cosmological problems

Sergey Siparov

State University of Civil Aviation, 38 Pilotov str., St-Petersburg, 196210;

Research Institute for Hyper Complex Systems in Geometry and Physics, 3 bg 1 Zavodskoy pr.,

Fryazino, Moscow region, 141190

Russian Federation

E-mail: sergey@siparov.ru

The gravitation theory is modified on the base of geometric identity and equivalence principle. This makes it possible to generalize the geodesics and solve several problems of classical GRT such as flat rotation curves of the spiral galaxies, Tully-Fisher law and some others and reveal the fundamental (geometrical) origin of the *cH* acceleration value. The developed approach contains all the results of the classical GRT and has promising cosmological consequences.

## 1. Introduction

The explanation of the flat character of the rotation curves of spiral galaxies is one of the most challenging problems of modern physics for the following reasons. It is a simple observable phenomenon easy to describe, it is not a small effect, it has more than satisfactory statistics, and for all that it seems to contradict the predictions of the pillar of modern astrophysics which is GRT, and Newton gravitation theory as well.

The attempts to modify the theory in order to describe the rotation curves have been undertaken for decades. The most tempting object for modification was the so called "simplest scalar" in the expression for the Hilbert-Einstein action

$$S_{EH} = -\frac{c^3}{16\pi G}\int d^4 x (-g)^{1/2} R \qquad (1.1)$$

Nobody knows the criterion for sufficient "simplicity", and in [1] the terms of the higher orders of scalar curvature were added, thus, giving birth to the so called *f(R)*-theories. One could also think of the use of an additional scalar field (still not found) like in [2] or of choosing a scalar originating from another rank four tensor as in [3] (excludes gravitational waves). Another trend is represented by the ideas given in [4] and by the series of papers beginning with [5]. In the first of them, the scalar-vector-tensor gravitation theory was introduced, and it includes the repulsive fifth force (with a specific fifth force charge) characterized by vector field. In the second, the phenomenological MOND was introduced, and it suggests either to modify Newton gravitation law or Newton dynamics law in such a way as to fit the observational data. Both these approaches give acceptable fits, but fail to provide a reliable physical idea grounding the chosen terms or functions.

The issue of modifying the theory with regard to the rotation curves is even more complicated by the whole set of restrictions stemming from the observational data and discussed in [6]. The restrictions include the demand for an explanation of Tully-Fisher law for the luminosity of spiral galaxies

$$L_{lum} \sim v_{orb}^4 \qquad (1.2)$$

and of the globular clusters problem. The last one has two sides. On the one hand, the globular clusters that don't belong to the galaxy plane obey the usual Einstein or Newton gravitation and, therefore, there is no need for the theory modification with regard to the motion in the direction orthogonal to the galaxy plane (anisotropy?). On the other hand, too many of them are known to be located in the vicinity of the galaxy center instead of spending most of their time on the periphery in accordance with second Kepler law. In [6] it was argued there that none of the known proposals suffices all of these restrictions. One could also mention the lensing effect which confirms GRT qualitatively but is sometimes 4-6 times larger than predicted.

The conclusion is that the needed modification of the theory demands anisotropy which seems natural for a rotating spiral. In GRT the anisotropy was in a sense discussed when the rotation of the central mass was regarded. Presumably, the best known is Lense-Thirring effect [7] for which the precession of the gyroscope (e.g. planet) in the field of the rotating star was calculated. The results of the latest measurements performed by Gravity Probe B [8] coincide with the corresponding predictions within the accuracy higher than 1%. The particular class of phenomena described by Lorentz type gravitational forces acting on a probe particle moving nearby the spinning point mass is known as gravitoelectromagnetism (GEM field) which is the next order correction to the Newton theory as it follows from GRT. The tiny effects of what is now called rotational frame-dragging alongside with linear frame-dragging and static mass increase were discussed also by Einstein in [9]. The Coriolis forces characteristic for rotating frame, present an example of inertial forces depending on velocity of the body and on the velocity of the reference frame. Equivalence principle recollected, one could think of the velocity dependent gravitational forces. But such simple frame can hardly be used directly to describe the gravitation field of the rotating galaxies because they can not be regarded as compact spinning mass.

In this paper the geometrical approach is used to obtain the equations containing the velocity dependence of gravitational forces [10]. In order to do this, the anisotropic metric is introduced with regard to the general geometric identity (known as Maxwell equation in mathematics) and to the equivalence principle. This metric could lead not to an arbitrary but to the natural change in the "simplest scalar". From the physical point of view, the velocity dependent gravitation forces are consistent, since we postulate the impossibility to distinguish between inertial forces and gravitational ones. The geometrical approach causes the appearance of the new fundamental constant similarly to the situation when Minkowski space-time was introduced. Then a fundamental *velocity* appeared in metric and found a physical interpretation in the relativity theory. Now the fundamental *angular velocity* appears, and it also finds physical interpretation.

## 2. Anisotropic perturbation and generalized geodesics

### 2.1 Metric and geodesics

In order to account for anisotropy in sources distribution in an object like a spiral galaxy, let us regard an anisotropic space with a deformed metric of the following form

$$\widetilde{g}_{ij}(x,u(x),y) \equiv g_{ij}(x,y) = \gamma_{ij} + \varepsilon_{ij}(x,y) \tag{2.1}$$

where $\gamma_{ij}$ is x-independent metric (here: Minkowski one), $\varepsilon_{ij}(x,y)$ is a small anisotropic perturbation, $y$ belongs to the tangent space, and along some curve (trajectory of the probe particle), $x^i = x^i(s)$ we shall always consider $y^i = \dfrac{dx^i}{ds}$, and, finally, $u(x)$ is the vector field corresponding to the motion of sources and it generates the anisotropy. Notice, that every point of the main manifold is supplied by *two* vectors belonging to a tangent space. The tangent bundle of a space with an anisotropic metric becomes an eight dimensional Riemannian manifold equivalent to the *phase space*. On this bundle, the local coordinates are $(x^i, y^i)$, where $x^i$ are positional variables, $y^i$ are the directional ones, and both must be treated in the same way (see Appendix). Euler-Lagrange equations can be obtained by varying the Lagrangian, $L = (\gamma_{hl} + \varepsilon_{hl}(x,y))y^h y^l$. In this case the expression for the generalized geodesics is obtained similarly to [12] and takes the form:

$$\frac{dy^i}{ds} + (\Gamma^i{}_{lk} + \frac{1}{2}\gamma^{it}\frac{\partial^2 \varepsilon_{kl}}{\partial x^j \partial y^t}y^j)y^k y^l = 0 \tag{2.2}$$

where $\Gamma^i{}_{jk} = \dfrac{1}{2}\gamma^{ih}(\dfrac{\partial \varepsilon_{hj}}{\partial x^k} + \dfrac{\partial \varepsilon_{hk}}{\partial x^j} - \dfrac{\partial \varepsilon_{jk}}{\partial x^h})$ is y-dependent Christoffel symbol.

*Remark:* The direction dependent metrics may define various geometries on anisotropic spaces. The most widely known is Finsler geometry [13] corresponding to Finsler metric tensor,

$g_{ij} = \frac{1}{2} \frac{\partial^2 F^2}{\partial y^i \partial y^j}$ where $F = F(x,y)$ is 1-homogeneous in $y$ and $det(g_{ij}) \neq 0$ for all $(x,y)$ on $TM$. But here we actually use a generalized Lagrange metric.

The generalized geodesics (2.2) will be used to follow the classical Einstein approach [11] step by step. Particularly, two of the simplifying assumptions we are going to use are just those introduced by Einstein when he derived Newton law, and the third assumption reproduces the second one with regard to the $y$-derivatives. This means that $\varepsilon(x,y)$ is again considered small enough to use a linear approximation.

The assumptions are the following:

1. The velocities of the material objects are much less than the fundamental velocity. This means that the components $y^2$, $y^3$ and $y^4$ can be neglected in comparison with $y^1$ which is equal to unity within the accuracy of the second order;

2. Since the velocities are small, the time derivative of metric $\frac{\partial \varepsilon_{hj}}{\partial x^1}$ can be neglected in comparison to the space derivatives $\frac{\partial \varepsilon_{hj}}{\partial x^\alpha}; \alpha = 2,3,4.$

3. The same is taken true for the $y$-derivatives: $\frac{\partial \varepsilon_{hj}}{\partial y^1}$ can be neglected in comparison to the space derivatives $\frac{\partial \varepsilon_{hj}}{\partial y^\alpha}; \alpha = 2,3,4.$

As in [11], the assumptions make it possible to preserve only the terms with $k = l = 1$, which means that the only $\varepsilon_{kl}$ remaining in the equation (2.2) is $\varepsilon_{11}$, while $y^k = y^l = 1$. Let us introduce the new notation for the $y$-derivative of the perturbation

$$\frac{1}{2} \frac{\partial \varepsilon_{11}}{\partial y^t} \equiv A_t \tag{2.3}$$

similar to a component of the Cartan tensor. Notice, that $A_t$ are the components of the $y$-gradient of $\varepsilon_{11}$, i.e. $A_\alpha = \frac{1}{2}(\nabla_{(y)} \varepsilon_{11})_\alpha$ for $\alpha = 2,3,4$ (the same numeration 1 to 4 is used for both $x$- and $y$-variables). Then we get

$$\frac{dy^i}{ds} + \Gamma^i{}_{11} + \gamma^{it} \frac{\partial A_t}{\partial x^j} y^j = 0 \tag{2.4}$$

The third term in the eq.(2.4) does not vanish since though we assume $\frac{\partial A_i}{\partial x^1} \ll \frac{\partial A_i}{\partial x^\alpha}$ but $y^1 \gg y^\alpha$ for $\alpha = 2,3,4$. Now add and subtract the same value $\gamma^{it} \frac{\partial A_j}{\partial x^t} y^j$ to obtain

$$\frac{dy^i}{ds} + \Gamma^i{}_{11} + \gamma^{it}[(\frac{\partial A_t}{\partial x^j} - \frac{\partial A_j}{\partial x^t}) + \frac{\partial A_j}{\partial x^t}]y^j = 0 \qquad (2.5)$$

The expression $(\frac{\partial A_t}{\partial x^j} - \frac{\partial A_j}{\partial x^t})$ can be taken as a component of an anti-symmetric tensor, $F_{jt}$, and eq.(2.5) yields for the generalized geodesics

$$\frac{dy^i}{ds} + \Gamma^i{}_{11} - \gamma^{it} F_{tj} y^j + \gamma^{it} \frac{\partial A_j}{\partial x^t} y^j = 0 \qquad (2.6)$$

For $\alpha, \beta = 2, 3, 4$ the expressions $(\frac{\partial A_t}{\partial x^j} - \frac{\partial A_j}{\partial x^t})$ are the components of the curl of vector $\vec{A}$.

If eq.(2.6) contained only two first terms, one would get the Einstein result [11]. If there were three first terms and an additional field with a 4-potential characterized by an interaction constant, $q$, one could think of an electromagnetic tensor and of electrodynamics. But no interaction but ineradicable gravitation was introduced.

## 2.2 The meaning of analogy with electromagnetism

It should be underlined that both in electromagnetic and gravitational cases the structures known as "Maxwell equations" can be deduced from a purely geometrical identity,

$$\frac{\partial F_{ij}}{\partial x^k} + \frac{\partial F_{jk}}{\partial x^i} + \frac{\partial F_{ki}}{\partial x^j} = 0 \qquad (2.7)$$

where $F_{ij}$ is an anti-symmetric tensor of the type mentioned above. It is only the historical tradition that could make one think that Maxwell equations are the generalization of solid physical data while GEM-like expressions are able only to give small second order Einstein corrections to Newton gravity.

Therefore, it seems worth to remind that rewriting eq.(2.7) explicitly and making a formal designation

$$F_{23} \equiv -B^{(g)}{}_z; F_{14} \equiv E^{(g)}{}_z; F_{31} \equiv E^{(g)}{}_y;$$
$$F_{24} \equiv B^{(g)}{}_y; F_{12} \equiv E^{(g)}{}_x; F_{34} \equiv -B^{(g)}{}_x \qquad (2.8)$$

one immediately obtains the homogeneous equations

$$\frac{\partial \vec{B}^{(g)}}{\partial t} + rot\vec{E}^{(g)} = 0$$
$$div\vec{B}^{(g)} = 0 \qquad (2.9)$$

Then, following the geometrical receipt [11], one may pass to the contra-variant tensor $F^{ij} = g^{ik} g^{jm} F_{mk}$, and introduce a new vector $I^i$ according to

$$I^i = \frac{\partial F^{ij}}{\partial x^j} \tag{2.10}$$

Notice, that if $F_{mk}$ is small and Minkowski metric $\gamma_{ik}$ can be used to get a linear approximation, $g_{ik} = \gamma_{ik} + \varepsilon_{ik}; \varepsilon_{ik} = \varepsilon \varsigma_{ik}; \varepsilon \ll 1$, then $\gamma^{it}$ can be used to raise the index. Making another formal designation,

$$I^1 \equiv \rho^{(m)}; I^2 \equiv j^{(m)}{}_x; I^3 \equiv j^{(m)}{}_y; I^4 \equiv j^{(m)}{}_z \tag{2.11}$$

one can obtain the inhomogeneous equations in a similar way

$$\begin{aligned} rot\vec{B}^{(g)} - \frac{\partial \vec{E}^{(g)}}{\partial t} &= \vec{j}^{(m)} \\ div\vec{E}^{(g)} &= \rho^{(m)} \end{aligned} \tag{2.12}$$

Regarding $\vec{E}^{(g)}$ and $\vec{B}^{(g)}$ as vortex-free and solenoidal parts of the vector part of the gravitational field, one gets $E^{(g)} = -\nabla(\int \frac{\rho^{(m)}(r)}{|r - r_0|} dV)$ and $B^{(g)} = rot(\int \frac{j^{(m)}(r)}{|r - r_0|} dV)$.

If we interpret density and current density in electric terms and introduce electric charge, $q$ to describe the interaction, we recognize Maxwell field equations and get Lorentz force in dynamics equations. But if we interpret density and current density in gravitation terms and introduce gravitation charge, i.e. gravitation mass, the situation changes, because according to the main postulate of GRT, i.e. the equivalence principle, the gravitational force that now corresponds to moving masses is not an external one but must *enter the metric*. This is the meaning of the corresponding term in eq.(2.6).

Turning back to eqs.(2.9, 2.12), one can see that according to the assumption $\frac{\partial A_i}{\partial x^1} \ll \frac{\partial A_i}{\partial x^\alpha}$ and to the definition (2.3), vector $\vec{E}^{(g)} = (F_{12}, F_{13}, F_{14})$ is equal to $E^{(g)} = -\nabla_{(x)} A_1$ where $A_1$ is the value of the first component of the y-gradient of $\varepsilon_{11}$, i.e. $A_1 = \frac{1}{2}(\nabla_{(y)}\varepsilon_{11})_1$. Vectors $\vec{E}^{(g)}$ and $\vec{B}^{(g)} \equiv rot_{(x)}\vec{A}$ were obtained out of the anisotropic metric and are related to the vector field $u(x)$ in the expression of metric. If it is possible to interpret $\vec{A}$ as vector potential of the gravitational field, $\rho^{(m)}$ as mass density of the source of gravity, and $\vec{j}^{(m)} = \rho^{(m)}\vec{V}$ as the density of the mass flow corresponding to the proper motion of the source and its parts, one obtains an impressive analogy with electromagnetism and all the formalism developed for it can be used in calculations. The discussion of the corresponding Einstein equations can be found, e.g. in [14].

## 3. Equation of motion and gravitation forces

Equation (2.6) resembles the geodesics given in [9] and presenting the next order approximation for $\frac{d^2 x^i}{dt^2} = \Gamma^i{}_{11}$ obtained in [11] for isotropic Riemannian space. In Einstein formula in [9], the lhs reflects the inertial mass increase when there are other masses nearby, the first term in the rhs corresponds to Newton gravity and the second and third terms in the rhs correspond to the rotational and linear frame-dragging effects. The expression similar to the second term in the rhs of [9] was also obtained and used in [7] and others for the additional acceleration produced by the spherical mass spinning with angular velocity $\Omega$. It has obvious relation to the Coriolis force.

In order to analyze eq.(2.6), notice first, that in the GEM-type force resulting from its second term, only the gravitomagnetic part remains while the gravitoelectric part of it can be neglected because of the third assumption. Then let us transform the third term in eq.(2.6) with regard to the second assumption and get

$$-\gamma^{\alpha t}\frac{\partial A_j}{\partial x^t}y^j = -(-\delta^\alpha{}_t)[\frac{\partial}{\partial x^t}(A_j y^j) - \frac{\partial y^j}{\partial x^t}] = (\nabla_{(x)}(A_j y^j))^\alpha = (\nabla_{(x)}(\vec{A},\vec{y}))^\alpha \quad (3.1)$$

where $\frac{\partial y^j}{\partial x^t}$ vanishes because $x$ and $y$ are independent variables. Recollecting that some terms in eq.(2.6) were initially multiplied by $y^1 y^1$ and $y^1 = 1$ unit of length, we introduce all the dimensional factors explicitly (see also Appendix) and get

$$H\frac{d\vec{y}}{c^2 dt} = \frac{1}{2}\{-\nabla_{(x)}\varepsilon_{11} + [\vec{y}, rot_{(x)}(\frac{\partial \varepsilon_{11}}{\partial y})] + \nabla_{(x)}((\frac{\partial \varepsilon_{11}}{\partial y}), \vec{y})\} \cdot (\frac{H}{c}y^1)^2 \quad (3.2)$$

Since $\vec{y} = (1/H)\vec{v}$ and $v^1 = c$, the expression for the gravitation force acting on a particle with mass, $m$, obtains the form

$$m\frac{d\vec{v}}{dt} = \vec{F}^{(g)} = \frac{mc^2}{2}\{-\nabla_{(x)}\varepsilon_{11} + [\vec{v}, rot_{(x)}\frac{\partial \varepsilon_{11}}{\partial \vec{v}}] + \nabla_{(x)}(\frac{\partial \varepsilon_{11}}{\partial \vec{v}}, \vec{v})\} \quad (3.3)$$

Let us mention some details concerning all the three terms in the figure brackets of the eq.(3.3).

The first term is related to the expression for the usual gravity force, $F^{(g)}{}_N$, acting on a particle with mass, $m$. For the stationary point source of gravitation with mass $M$, the solution of Poisson equation suggests $\varepsilon_{11} \sim 1/r$, where $r$ is the distance from the particle to the source, and in this case the expression $\varepsilon_{11} = \frac{2GM}{c^2}\frac{1}{r}$ in eq.(3.3) for the point source at sufficient distances would give Newton law $F^{(g)}{}_N = G\dfrac{Mm}{r^2}$. The value $r_S = \dfrac{2GM}{c^2}$ corresponds to Schwarzschild radius. This result will remain the same if the particle is at the periphery of the distribution of masses and *M* is an integral of mass density.

The second term can be recognized as related to Coriolis force which is proportional to velocity, $\vec{v}$ of the particle whose dynamics is described by eq. (3.3) and to the proper motion of the gravitation sources described by $rot_{(x)}(\dfrac{\partial \varepsilon_{11}}{\partial y})$. Introducing the notation

$$\Omega = \dfrac{c^2}{4} rot_{(x)} \dfrac{\partial \vec{\varepsilon}_{11}}{\partial \vec{v}} \qquad (3.4)$$

($\Omega$ may now depend on *x*), one gets the exact pattern of the Coriolis force

$$F^{(g)}{}_C = 2m[\vec{v},\vec{\Omega}] \qquad (3.5)$$

Thus, the actions produced by $F^{(g)}{}_C$ on a body could be attraction, repulsion and tangent action depending on the angle between $\vec{v}$ and $\vec{\Omega}$. The component of velocity, $\vec{v}$, which is parallel to $\vec{\Omega}$ is not affected by the second term in eq.(3.3), and this corresponds to one of the features of the globular clusters behavior mentioned in the Introduction.

Introducing specific vectors $\vec{\beta} = \vec{v}/c$ and $\vec{\Theta} = \vec{\Omega}/H$, in which *c* and *H* represent the geometrically motivated constants mentioned in Appendix, one obtains

$$F^{(g)}{}_C = 2m \cdot cH \cdot [\vec{\beta},\vec{\Theta}] \qquad (3.6)$$

If we interpret the (geometrical) fundamental velocity, *c*, as the speed of light (as it is usually done) and the (geometrical) measurement units factor *H* as Hubble constant, we find out that the origin of the value of numerical factor which was noticed and discussed many times in astrophysics and gravitation theory modifications stems from geometry. When the product *βΘ* approaches unity the value of additional acceleration approaches *cH*.

The third term corresponds to the action produced on a moving particle by radial expansion (explosion) or by radial contraction (collapse) of the system of gravitating sources. The particle suffers an additional attraction to or repulsion from the center of mass distribution depending on the sign of scalar product. If the system of sources expands and the particle moves radially inwards, or if the system of sources contracts and the particle moves radially outwards, there is an additional attraction. If the system of sources expands and the particle moves radially outwards, or if the system of sources contracts and the particle moves radially inwards, the particle suffers a repulsion from the center of mass distribution.

Thus, the characteristic features of the anisotropic geometrodynamics (AGD) approach are the following. The total acceleration of the probe particle can now depend not only on the location of distributed masses but also on their proper motion and on the motion of the particle itself. Notice, that in AGD the gravitational interaction ceases to be simple attraction as before, it depends on the motion of the particle and of the sources and can be attraction, repulsion and transversal action. The value of *cH* which earlier had an empirical origin may now be regarded as an intrinsic (geometrical) property of the theory. It goes without saying that all the GRT results remain valid for a planetary system scale.

## 4. AGD applications

Due to the character of the theory developed here, now there is no need for the concrete observations data to fit for. But we certainly have to make sure that the qualitative picture is correct.

The spiral galaxies have natural preferential direction. In order to get the results comparable with observations, let us introduce a simplified model and discuss its properties. Let a system consist of a central mass and an effective circular mass current, $J^{(m)}$ around it. For galaxies like M-104 (Sombrero) or NGC-7742 with the pronounced ring structure this model can be used at once, for other galaxies – with emphasized spiral arms – the effective values of contour radius, $R_{eff}$, constant angular velocity, $\Omega_{eff}$, and linear velocity of mass density motion along the contour, $V_{eff} = \Omega_{eff} R_{eff}$, should be introduced. It could be done, for example, in the following way

$$I_{eff} = \sum I_n \equiv M R_{eff}^2 \Rightarrow R_{eff}^2 = \sqrt{\frac{I_{eff}}{M}} \tag{4.1}$$

where $I_{eff}$ is the moment of inertia of the system with the total mass, $M$. The effective angular velocity, $\Omega_{eff}$, can be defined from $I_{eff} \Omega_{eff} \equiv L_{eff} = \sum L_n$, where $L_n$ is the angular momentum of the component of the system. We get, thus,

$$\Omega_{eff} = \frac{L_{eff}}{I_{eff}} \tag{4.2}$$

These parameters can be estimated for a chosen galaxy from the astronomical observations.

This model clarifies the reason for the non-Keplerian behavior of globular clusters: their motion is affected not only by the gravitation center but by the effective contour too (see subsection 4.6). If we consider mass distribution in a spiral galaxy as a whole to be radially stationary (at least in comparison with orbital motion), the third term in eq.(3.3) can be neglected.

Due to the identity of the origin of Maxwell equations for electrodynamics and for gravitation mentioned above, such model is quite similar to electromagnetic one with a charge at the center and a circular electric current around it, thus, the mathematical results from electrodynamics can be used in calculations dealing with velocity dependent gravitation.

*4.1 Flat rotation curve*

In order to describe the spiral galaxy with a bulge, let us use the electromagnetic version of the model and regard a positive charge, a circular contour with current, $J$ around it and an electron orbiting the system in the plane of the contour. Strictly speaking, an electron in such a system can not be in a finite motion and has either to fly away or to fall on the center. This provides an idea of the arms origin in spiral galaxies which is mentioned in subsection 4.6. But the number of electron rotations could be large enough. The value of $B_z(r)$ component of the magnetic induction produced by the contour with radius, $R_{eff}$, can be found with the help of Bio-Savart law and according to [15] with $c = 1$ is equal to

$$B_z(r) = J \frac{2}{\sqrt{(R_{eff}+r)^2 + z^2}} [K + \frac{R_{eff}^2 - r^2 - z^2}{(R_{eff}-r)^2 + z^2} E]$$

$$K = \int_0^{\pi/2} \frac{d\theta}{\sqrt{1-k^2 \sin^2\theta}} ; E = \int_0^{\pi/2} \sqrt{1-k^2 \sin^2\theta}\, d\theta \qquad (4.3)$$

$$k^2 = \frac{4 R_{eff} r}{(R_{eff}+r)^2 + z^2}$$

where $K$ and $E$ are the elliptic integrals. Introducing notation, $b = r/R_{eff}$, and taking $z = 0$, one gets

$$B_z(r) = J \frac{2}{R_{eff}(1+b)} [K + \frac{1-b^2}{(1-b)^2} E] \quad (4.4)$$

The internal region close to the charge corresponds to $b \ll 1$ and $B_z(r) \to J/2R_{eff}$, the far away region corresponds to $b \gg 1$ and $B_z(r) \to 0$, and the intermediate region to which the contour also belongs corresponds to $b = O(1)$ and

$$B_z(r) \sim J/r \qquad (4.5)$$

The centrifugal force acting on the orbiting electron, $m\frac{v_{orb}^2}{r}$ is equal to the sum of the Coulomb attraction, $F_{Cl} = qC_1/r^2$, produced by the central charge and the Lorentz force, $F = qv_{orb}B_z(r)$. For the intermediate region corresponding to the periphery of a galaxy, the dynamics equation with $J \equiv C_2$ can be written as

$$mv_{orb}^2 = \frac{qC_1}{r} \pm qv_{orb}C_2 \qquad (4.6)$$

where $C_1$ and $C_2$ are constants characterizing the system and the sign corresponds to the direction of current and the location of the electron inside or outside the contour. Applying the result to the gravitational case, we have to substitute the electric charge by the gravitational one, $q = m_g$, and use the equivalence principle, $m_g = m$. The smaller root of the square equation (4.6), $v_{orb} = \frac{C_2}{2}(1 - \sqrt{1 \pm \frac{4C_1}{rC_2^2}})$ corresponds to Newton law, the sign depends on the direction of the electron motion. Neglecting the small term inside the square root in the larger root of the equation, $v_{orb} = \frac{C_2}{2}(1 + \sqrt{1 \pm \frac{4C_1}{rC_2^2}})$ one gets

$$v_{orb} \sim C_2, \qquad (4.7)$$

which corresponds to the flat rotation curve on the periphery.

### *4.2 Tully-Fisher law*

Let us estimate $C_2(R_{eff}) = J^{(m)}(R_{eff})$. The mass current is given by $J^{(m)}(R_{eff}) \sim M/T$ with $M$ proportional to the area of a spiral galaxy, $R_{eff}^2$, and the period $T \sim R_{eff}^{3/2}$ according to Kepler law. This gives $J^{(m)}(R_{eff}) \sim \sqrt{R_{eff}}$. Since the luminosity, $L_{lum}$, is also proportional to the galaxy area, we get $R_{eff} \sim \sqrt{L_{lum}}$. Therefore, $J^{(m)}(R_{eff}) \sim \sqrt{R_{eff}} \sim L_{lum}^{1/4}$ and

$$v_{orb} \sim L_{lum}^{1/4} \qquad (4.8)$$

which corresponds to the Tully-Fisher law (1.2).

### *4.3 Applicability region*

The results presented by eq.(4.7) and eq.(4.8) suggest to estimate the regions and regimes for which this or that term in eq.(3.3) plays an essential role. With regard to definitions and eqs.(3.4, 3.5), we can take $\Omega \sim \frac{G}{c^2 r}\frac{MV_{eff}}{R_{eff}}$ and the acceleration given by $a_C = 2[\vec{v}, \vec{\Omega}]$ is proportional to

$$a_C \sim v\frac{GMV_{eff}}{c^2 rR_{eff}} \qquad (4.9)$$

Then the ratio of this acceleration to the Newtonian one, $a_N = \dfrac{GM}{r^2}$ is

$$\frac{a_C}{a_N} \sim \frac{vV_{eff}}{c^2} \frac{r}{R_{eff}} = \frac{vr}{c^2} \Omega_{eff} \qquad (4.10)$$

and one can estimate the region where the specific features of AGD become significant. For a given particle moving with velocity, *v,* at the distance, *r*, from the center of spiral galaxy this ratio becomes

$$\frac{a_C}{a_N} \sim \frac{vr}{c^2} \frac{L_{eff}}{I_{eff}} \qquad (4.11)$$

where $L_{eff}$ and $I_{eff}$ characterize the galaxy. Every concrete case must be considered with regard to eq. (4.11).

### *4.4 Giant black hole in the center of a spiral galaxy*

Let us regard an illustrative limit case when the contour with mass current is close to the rim of the giant black hole in the center of a galaxy. Its mass is *M*, the effective radius is $R_{eff} = r_S = \dfrac{2GM}{c^2}$ and the effective velocity is equal to orbital velocity $V_{eff} = c$. Then the acceleration ratio eq.(4.10) will be equal to

$$\frac{a_C}{a_N} \sim \frac{vV_{eff}}{c^2} \frac{r}{R_{eff}} = \frac{cvr}{2GM} \qquad (4.12)$$

Let us find the distance at which both accelerations are equal and $M = 10^{11}$ Solar masses (e.g. as in Milky Way)

$$1 \sim \frac{cvr}{M_{Sol}} \sim 10^{-23} vr \qquad (4.13)$$

We see that the measured observable orbital velocity of stars at the periphery, $v \sim 10^5 m/s$, corresponds to the distance, $r \sim 10^{18} m$, which is in accord with the estimation for the galaxy radius. This means that there is no reason to expect the behavior of the rotation curves to be Newtonian.

### *4.5 Pioneer anomaly*

The so called Pioneer anomaly presents the existence of the measured extra sunward acceleration of the probes Pioneer 10 and Pioneer 11 equal to *(8.74 ± 1.33)·$10^{-10}$m/s* [16] which for the distance *r ~ 68a.u.* from the Sun makes 0.065% of the Newtonian term. In frames of our approach the extra gravity could be caused by the rotational motion of the planets and of the Sun itself. The qualitative estimation of the possible additional deceleration corresponding to the Solar system including the Sun and the planets can be performed with the help of eq. (4.11). For the probe velocity taken as $v \sim 4 \cdot 10^4 m/s$, we get an extra deceleration equal to 0.0064% of the Newtonian term and see that qualitatively the effect takes place but it is an order less than needed. The cause of it might be the wrong choice of the theory for a given scale - similarly to the case when the classical GRT is used to describe the galaxies. In AGD an essential role is played by the motion of distributed masses, but in a planetary system the total mass of the planets is negligible in comparison with the mass of the rotating Sun. Notice that the GEM-theory gives the same (insufficient) order of the additional acceleration, $2.1 \cdot 10^{-11} m/s^2$, suffered by a probe particle at the Earth orbit radius under the action of Sun rotation.

### 4.6 Numerical modeling

The simplified center plus current (CPC) model makes it possible to obtain some visual results with the help of numerical calculations.

On Fig.1 one can see various regimes of motion that depend on the parameters of the system and on the initial conditions. To the left is the quasi-precession of an orbit in the CPC

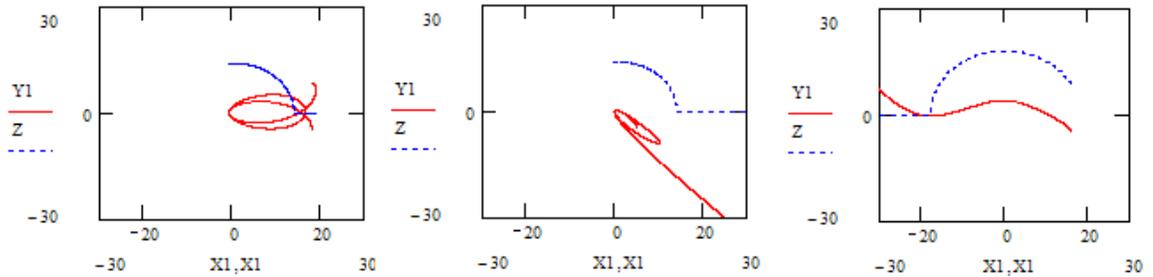

Fig.1. AGD based trajectories mimicking various observable phenomena: left – quasi-precession; middle – non-Keplerian behavior of globular clusters; right – double bending flyby

system; in the center there is a trajectory illustrating the tendency of a globular cluster to be present in the vicinity of the center longer than it should according to Kepler; and to the right is the flyby of a particle through a CPC system presenting two bends. The graph in the right also relates to the problem of gravitational lensing. In the AGD, the lensing effect predicted by GRT might be not only amplified by the effective mass current, but also be attenuated and even obtain the opposite sign depending on initial conditions as shown on the figure. The negative lenses (like the one shown on the figure) diminish the angular size of the objects behind them.

Therefore, if there is such a lens between an astronomical standard candle and the observer, the distance to it might be considered larger than it really is. This could be the reason of the interpretation of the recent supernovas 1a observations as pointing at the acceleration of the Universe expansion.

More than 50% of all known galaxies are spiral, the forms of their arms essentially vary and two thirds of all the spirals have bars. As it is known in astronomy, the bulge of a spiral galaxy consists of old stars while the arms consist of young stars. Alongside with density wave theory of the arms' origination (which doesn't explain the flat rotation curves), one may think of two other possibilities concerning the evolution of a spiral galaxy. Either young stars are formed far from the center, move towards it along similar trajectories, attracting each other to form noticeable arms and finally get old and disappear. Or the bulge is an active zone that produces stars; they are thrown away by huge explosions and then are involved into galaxy rotation. In both cases it is hard to explain the form of arms using only the known theory.

On Fig.2 one can compare the Hubble image of NGC-1365 galaxy to the trajectories built for the two bodies with symmetrical initial conditions moving towards (central graph) and outwards (right graph) the center of a galaxy according to the AGD based model. The exact form of the little features in the center of the right graph depends on the step of calculations, but the general pattern remains the same.

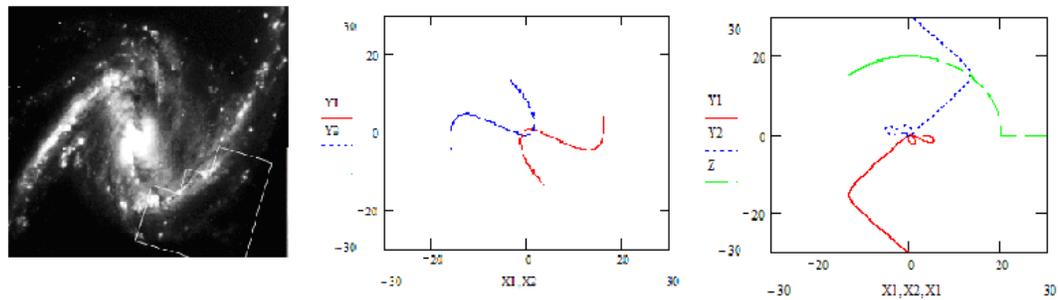

Fig.2. Hubble image of NGC-1365 galaxy and symmetrical trajectories

**5. Discussion**

The geometrical identity leads to Maxwell equations independently of the origin of physical field that could be used for interpretation. Therefore, the proper motion of the distributed sources could affect the gravitational phenomena as well as electromagnetic ones. The difference is that the equivalence principle demands to account for this action not by introduction of an extra (gravitomagnetic type) force, but by the modification of the space-time metric which becomes anisotropic. This anisotropy is interpreted as the dependence of gravitation forces acting between the bodies not only on their position but also on their motion. The proper motion of the distributed sources of gravitation adds extra terms to the gravitation field, and the moving probe body interacts with it with account to its own velocity as shows eq.(3.3).

It seems clear that all the results of the GRT remain valid when the proper motion of the sources can be neglected. But when many gravitation sources start to move relative to each other and the scale of phenomena in question grows, the uncovered anisotropy in geometrodynamics starts to play an essential role as shown by eqs.(4.10, 4.11). This is demonstrated by the explanation of the flat character of the rotation curves of spiral galaxies, eq.(4.7), which can't be done in frames of classical GRT. It turns out that the AGD approach also explains the empirical Tully-Fisher law, eq.(4.8), provides the fundamental (geometrical) origin to the *cH* acceleration value, eq.(3.6), and suffices the astrophysical restrictions for the gravitation theory modifications with concern to the observed motion of globular clusters. The role of the third term in the expression for the gravitation force eq.(3.3) might appear important for radial instabilities of mass distributions like explosions and collapses. The analogy with electromagnetism suggests a wide variety of possible phenomena to explain and look for. For example, the possible existence of negative gravitational lenses could be used for the interpretation of the supernovas 1a observations with no acceleration of the Universe expansion; an AGD based simple model provides the recognizable form of arms of the spiral galaxy with a bar.

These ideas have far going perspectives in cosmology in general where the following new direction of thought appears. First, we see that the flat rotation curves can be explained without introduction of *dark matter* notion for galaxies, and it makes one think that the gravitational binding in galaxy clusters could be also provided by additional gravitational force due to the relative motion of galaxies. Second, according to AGD, the repulsive forces acting on the cosmological scale could be provided by the velocity dependent gravitation, therefore, the notion of *dark energy* of repulsion starts to cause specific doubts. Third, the Hubble red shift the explanation of which now refers to the (infinite) Universe expansion could be also caused by the tangent motion of huge masses on the periphery of the (finite) visible part of the Universe – such motion would cause the additional gravitation force and, consequently, the *gravitational* red shift with accord to fundamental GRT ideas. This suggestion finds support in the observations of the tangent motions of distant quasars – they take place at amazingly high velocities [17]. Finally, the AGD provides a new insight for Mach's principle and for the border of the Universe problem.

### 7. Acknowledgments

I would like to express my gratitude to N. Brinzei, S.Kokarev and the anonymous A&A referee for the helpful discussions and criticisms.

This work was supported by the RFBR grant No. 07-01-91681-RA_a.

## 8. Appendix

On the tangent bundle $x^i$ is regarded as a positional variable, and $y^i$ is a directional variable and is proportional to

$$y^i \sim \frac{dx^i}{ds}, \tag{A1}$$

where $s$ is a parameter usually taken as an arc length. Since on this 8-dimensional manifold $x$ and $y$ have to be treated in a similar way, there must be a dimensional factor in the definition of $y^i$ chosen such that the measurement units of $y^i$ are the same as those of $x^i$

$$y^i = l\frac{dx^i}{ds}; [l] = length \tag{A2}$$

Such definition makes it possible to have the simplest case of the Sasaki lift [18], i.e. to use the same metric tensor to raise and lower indices in both $x$ and $y$ subspaces of the tangent bundle. When we turn to physical problems, it is convenient to use time, $t$, instead of an arc length, $ds = cdt$, where $c$ is a constant with the dimensionality of speed, $[c] = distance/time$. Then

$$y^i = l\frac{dx^i}{cdt} = \frac{1}{H}\frac{dx^i}{dt} = \frac{1}{H}v^i; [H] = (time)^{-1} \tag{A3}$$